\documentclass[%
 aip,
 amsmath,amssymb,
 reprint,
 floatfix,
 nofootinbib
]{revtex4-1}

\usepackage{graphicx}
\usepackage{dcolumn}
\usepackage{bm}
\usepackage[utf8]{inputenc}
\usepackage[T1]{fontenc}
\usepackage{mathptmx}
\usepackage{etoolbox}
\usepackage{siunitx}
\usepackage{booktabs}
\usepackage{multirow}
\usepackage{xcolor}
\usepackage{hyperref}

\usepackage[version=4]{mhchem}
\usepackage{array}
\usepackage{tabularx}
\usepackage{gensymb}

\graphicspath{{figures/}}

\DeclareMathAlphabet{\mathcal}{OMS}{cmsy}{m}{n}

\makeatletter
\let\NAT@citesuper\NAT@cite
\makeatother

\setcitestyle{numbers,square,comma,sort&compress}

\newcolumntype{Y}{>{\centering\arraybackslash}X}

\begin{document}

\raggedbottom

\title[VE-DeOh Stringiness of Unentangled Polymer Solutions]{Viscoelastocapillary Extensional \emph{De-Oh} or \emph{VE-DeOh} Stringiness of Unentangled Polymer Solutions}

\author{Louis Edaño}
\affiliation{Department of Chemical Engineering, University of Illinois Chicago, 929 W Taylor St., Chicago, Illinois 60607, USA}

\author{Vivek Sharma}
\email{viveks@uic.edu}
\affiliation{Department of Chemical Engineering, University of Illinois Chicago, 929 W Taylor St., Chicago, Illinois 60607, USA}

\date{\today}

\begin{abstract}
Characterization, control, and calibration of stringiness, the propensity to form long, thin, persistent threads, are key to the design and application of polymer solutions and formulations transferred to substrates by dropwise dispensing, jetting, spraying, or coating, and are used in manufacturing fibers, membranes, or spray-dried products. For Newtonian fluids, enhancing shear viscosity translates into increasing stringiness as can be easily perceived during dripping of water vs. sugar syrups or glycerol water mixtures. In contrast, even polymer solutions with comparable shear viscosity can display a significant contrast in extensional rheology and apparent stringiness. However, there are no quantitative maps, formulas, or scales for stringiness, which motivates this study. In this contribution, we contrast the stringiness for a series of aqueous polymer solutions by determining the filament lifespan using dripping-onto-substrate (DoS) rheometry and filament length and lifespan in using dripping-into-air (DiA). We investigate the influence of polymer chemistry and molecular weight on stringiness and rely on DoS rheometry to characterize pinching dynamics and extensional rheology response. We show that an increase in stringiness correlates with steady, terminal extensional viscosity, and extensional relaxation time, respectively. Lastly, we provide a framework, which compares the stringiness of different polymer solutions through an extensional Ohnesorge-Deborah ($Oh-De$) plot or $Oh_E-De_E$ map, by computing the two dimensionless groups $Oh_E$ using steady, terminal extensional viscosity and $De_E$ using extensional relaxation time, respectively.
\end{abstract}

\maketitle

\section {Introduction}
Stringiness of liquids, including polymer solutions, often refers to their propensity to form long, thin, persistent filaments. It is a heuristic sensory or visual property that is apparent when stretching a finite liquid volume between two substrates, such as a finger and a thumb (a digital rheometer), between two metal plates, or between a falling drop and a stationary nozzle \cite{mckinley_visco-elasto-capillary_2005, petrie_extensional_2006,jones_stringiness_1982}. The length and longevity of the stretched liquid filament formed as a fry, glass or metal cylinder, finger, or spoon is pulled out and away from a pool is often used to judge the pick, cohesion, tackiness, ropiness, and/or stringiness of cosmetic products and coating fluids \cite{mckinley_visco-elasto-capillary_2005, petrie_extensional_2006,jones_stringiness_1982, tirtaatmadja_filament_1993, tirtaatmadja_drop_2006}. Many food aficionados check the stringiness and thickness of syrups and sauces by dripping from a ladle or stretching between a spoon and thumb \cite{mckinley_visco-elasto-capillary_2005, al_zahabi_pinching_2024,jones_stringiness_1982}. The suitability of a fiber spinning dope is often judged by stretching a filament between a pool and a rod withdrawn after a partial pool-dip, implying that stringiness anticipates spinnability. In agricultural sprays, pharmaceutical sprays, print inks, and coatings, the addition of relatively minute amounts of polymer diminishes the formation of satellite drops, allowing for tunability of drop size distribution in jetting, spraying, and coating processes \cite{ardekani_dynamics_2010, clasen_how_2006, bhat_formation_2010, sharma_rheology_2015, keshavarz_studying_2015, martinez_narvaez_dynamics_2021, jimenez_capillary_2020}. Stringiness appears to correlate with the sensory perception of the texture of foods \cite{jimenez_capillary_2020}, the cohesive and bonding ability of adhesives, and the uniformity and consistency of coatings. The preferences of a sensory panel for dripping and dipping food, pharmaceutical, cosmetic, and personal care products are influenced by stringiness \cite{hallmark_characterization_2016, al_zahabi_pinching_2024, jimenez_capillary_2020, mckinley_visco-elasto-capillary_2005, jones_stringiness_1982, tirtaatmadja_filament_1993, tirtaatmadja_drop_2006}. Furthermore, many biofluids such as okra water, deadly sundew mucilage, and life-affirming saliva and synovial fluids display stringiness, often attributed to the influence of dissolved biomacromolecules like polysaccharides and polypeptides \cite{al_zahabi_pinching_2024, bhat_formation_2010, mckinley_visco-elasto-capillary_2005, gaume_viscoelastic_2007, yuan_extensional_2018}.  Despite the importance attached to stringiness in biology and in industrial and everyday processes like dripping, dipping, spraying, splashing, printing, and coating, there is a lack of clarity about measuring stringiness, its relationship to shear and extensional rheology, and eventually to macromolecular properties, inspiring this investigation.

The saga of stringiness and spinnability, and their primary dependence on extensional or elongational viscosity, involves several key plot developments summarized by Petrie in several reviews \cite{petrie_extensional_2006}. Pioneering experiments and theoretical analysis by Trouton \cite{trouton_coefficient_1906} in 1906 considered the resistance to extensional or elongational flows arising for a cylinder in traction or axial compression, for free-falling jet-stream and sagging of a horizontal beam. Trouton showed that Netwonian fluids show an extensional viscosity, $\eta_E$ ("the coefficient of viscous traction"), that is a factor three times greater than the shear viscosity. To honor his contributions, we say that Newtonian fluids display a Trouton ratio of three or $Tr=\eta_E/\eta=3$. In 1908, Fano \cite{fano_arch_1908} compiled his study of thread-forming materials ("i corpi filanti"), including examples of many biological complex fluids with remarkable stringiness that he attributed to the role of macromolecules (polysaccharides and proteins). Fano also noted the possibility of a "tubeless siphon" and reported on elasticity of the thread-forming materials \cite{fano_arch_1908, petrie_extensional_2006, petrie_one_2006}. Scott Blair \cite{blair1941variations} and Clift \cite{clift1945observations} associated flow elasticity with spinnability, later Zidan \cite{zidan1969rheologie} found examples where this was not the case. Within half-a-century of Trouton and Fano's experiments, terms like "Spinnbarkeit", "Fadenziehens" "fibrosity" and spinnability were used in descriptions of long-lived filaments observed in experiments with printable inks, fiber spinning dopes, plant extracts, and various industrial fluids \cite{fano_arch_1908, blair1941variations, erbring_uber_1936, clift1945observations, petrie_extensional_2006, petrie_one_2006}. Ziabicki, Taksermann-Krozer and others studied stringiness using dripping or stretching liquid bridges in experiments often aimed at creating a better understanding of fiber spinning, processability or sensory response \cite{fano_arch_1908, ziabicki_mechanism_1964,tammann_uber_1927}. Many bespoke experimental methods were developed before the 1980s to measure the response to extensional flows, and were often used to measure spinnability, stringiness, ropiness, or Spinnbarkeit \cite{larson_spinnability_1983, sridhar_overview_1990}. In the late 1980s during the M1 Project, a cross-laboratory comparison of apparent extensional viscosity data for the same polyisobutylene-polybutene solution (M1 fluid) demonstrated that the datasets showed considerable variability, later attributed to deformation history and transient effects, and limitation on rates or strains that could be acquired \cite{sridhar_overview_1990, james_critical_1993}. Significant progress in experimental \cite{tirtaatmadja_filament_1993, bazilevskii_breakup_2001, stelter_validation_2000, stelter_investigation_2002, tirtaatmadja_drop_2006, sharma_rheology_2015, clasen_beads--string_2006, clasen_dispensing_2012, clasen_how_2006, vadillo_evaluation_2010, vadillo_microsecond_2012, dinic_extensional_2015}, theoretical, and computational \cite{yarin_free_1993, entov_effect_1997, prabhakar_effect_2017, prakash_universal_2019, wagner_analytic_2015, fontelos_evolution_2004} methods since this project has presented a better understanding of the connection between macromolecular properties, extensional viscosity, and their influence on drop or fiber formation \cite{tirtaatmadja_drop_2006, mckinley_visco-elasto-capillary_2005}. Nevertheless Petrie's reviews \cite{petrie_extensional_1995, petrie_extensional_2006}, especially from 2006, concluded that despite its significance, there is no agreed quantitative measure of spinnability. 

Innumerable papers in the last twenty years or so, especially on electrospinning and more recently on centrifugal spinning, correlate spinnability  with the presence of entanglements, assessed on the basis of success in producing fibers \cite{shenoy_role_2005, gupta_electrospinning_2005, haward_shear_2012, merchiers_evaporation_2021, badrossamay_nanofiber_2010, merchiers_fiber_2022, merchiers2022extensibility}. The entanglement concentration, $c_e$ is often inferred from shear viscosity measurements, for specific viscosity \cite{shenoy_role_2005, gupta_electrospinning_2005, haward_shear_2012} shows a strong power law increase with concentration (exponent $\eta_{sp}\sim c^{\alpha}; \alpha>3.9$). However, not all entangled polymer solutions result in fiber formation \cite{merchiers_evaporation_2021, merchiers2022extensibility}, as the processing conditions that lead to solidification must also be considered. It is well-known that viscous and polymer-free glass can form fibers, and that cotton candy is made from sugar syrup. Furthermore, many unentangled solutions can be fiber-spun by incorporating a high molecular weight polymer additive \cite{merchiers_evaporation_2021, merchiers_fiber_2022, merchiers2022extensibility,fang_manipulating_2015}, or through pathways to solidification, including by photopolymerization \cite{banerji2019cross, kim2019degradable, perry2024effect, liu2020effect}. Due to the influence of the fiber spinning process on the ability to form fibers, Larson \cite{larson_spinnability_1983} proposed intrinsic spinnability as a property dependent on the material response that is correlated with the ability to draw it into filaments or threads, but considers cases where surface tension does not play a role. In our recent studies \cite{merchiers_evaporation_2021, merchiers_fiber_2022}, we decided to reserve the term spinnability exclusively for cases where fiber spinning is an end goal, and refer to the material response consistent with the ability to make long, thin persistant filaments as stringiness. For unentangled polymer solutions, we wish to define viscoelastocapillary stringiness or extensional stringiness as the propensity to form long, thin persistent filaments by stretching or dispensing liquid bridges. Thus we seek a definition based primarily on material properties, probably captured historically in terms ranging from Spinnbarkeit, fibrosity and intrinsic spinnability.

Capillarity-driven or free-surface flows underlie heuristic estimates of stringiness made by dripping (into air or onto a substrate), by watching the filament formed on withdrawing a dipped rod or spoon, or by stretching a liquid bridge between two substrates. In liquid necks formed on dripping, jetting, or stretching liquid bridges, capillary stress seeks to strangle the pinched neck or break a cylindrical filament into drops. For polymeric solutions and formulations, macromolecular deformation, stretching, and orientation, contribute an extra resistance to the pinching and thinning that liquid necks experience in extensional flows \cite{entov_effect_1997, yarin_free_1993, anna_elasto-capillary_2001, wagner_analytic_2015, tirtaatmadja_filament_1993, dinic_rheology_2022, prabhakar_effect_2017, haward2013extensional, yokokoji2024rheological}. The pinching dynamics of Newtonian fluids are influenced by the interplay of capillary, viscous, inertial stresses, plus the tensile force in stretched necks. Non-Newtonian fluids involve accounting for additional contributions from elasticity or non-Newtonian viscosity \cite{mckinley_how_2000, clasen_how_2006, yarin_free_1993, wagner_analytic_2015}. The flow field in a pinching neck is extensional, and therefore the viscous stress for a Newtonian fluid is computed by recalling that its extensional viscosity, $\eta_E=3\eta$, is constant and three times the shear viscosity. Additional stresses arise for non-Newtonian fluids that can be written for polymer solutions by taking into account chain dynamics, flow-induced conformational changes, hydrodynamic and thermodynamic interactions, and entropic elasticity \cite{mckinley_visco-elasto-capillary_2005, yarin_free_1993, prabhakar_effect_2017, dinic_flexibility_2020}. The approach requires the use of a model, such as Oldroyd-B, Giesekus, FENE-P, PTT, etc. as summarized in various studies \cite{mckinley_visco-elasto-capillary_2005, yarin_free_1993, prabhakar_effect_2017, boyko_perspective_2024, ardekani_dynamics_2010}. The response of macromolecules to extensional flows within pinching necks influences the neck shape and radius evolution and therefore sets the apparent stringiness \cite{mckinley_visco-elasto-capillary_2005, entov_effect_1997, yarin_free_1993, clasen_how_2006, petrie_extensional_2006, dinic_extensional_2015, dinic_macromolecular_2019}. The contributions by polymers to the outcomes in capillarity-driven flows are assessed and analyzed in terms of the influence on elasticity (inferred as an extensional relaxation time) and transient or steady measures of extensional viscosity, as summarized next.   

The metrics for characterizing stringiness include filament length, as well as filament lifespan or persistence of a filament over time, though in many cases, the enhanced $\eta_E$ is correlated with increased stringiness. Filament length can be obtained by stretching liquid volume at a constant rate or a constant velocity, or by measuring the length of a filament formed on dripping or on withdrawal after dip-coating and this can be accomplished in a few different ways. Jones and Rees \cite{jones_stringiness_1982} in 1982 used dripping experiments to investigate the stringiness of dilute polymer solutions, and noted that an increase in nozzle size leads to larger drops, resulting in stronger tensile force in liquid necks and larger extensional rate. Matta and Tytus in 1990 \cite{matta_liquid_1990} stretched a liquid bridge between a fixed and a freely-falling cylinder and analyzed neck thinning rates under constant tensile force due to weight to obtain extensional viscosity measurements. In 1991, Sridhar and coworkers pioneered the filament stretching extensional rheometer (FISER) that measures the force or stress required for stretching a liquid bridge at a constant extension rate, enabling measurement of extensional viscosity, which correlates to stringiness \cite{sridhar_measurement_1991, tirtaatmadja_filament_1993,mckinley_filament-stretching_2002}. However, the FISER measurements are commonly restricted to low deformation rates and rather high viscosity polymer solutions and melts \cite{sridhar_measurement_1991, tirtaatmadja_filament_1993, mckinley_filament-stretching_2002, gupta2000extensional, mckinley_visco-elasto-capillary_2005}. Continuous filament stretching measurements with Trimaster and VADER have explored higher rates and lower viscosity systems than accessible with FISER \cite{vadillo_microsecond_2012, valette_effect_2019, aisling2024importance}. The measurement of force required to stretch a fluid between metal plates at constant speed is the basis of a tack test carried out on a Texture Analyzer or on a rheometer with a suitable normal force transducer. The tack test is often used in industrial practice to compare stringiness or ropiness, or cohesiveness of food and cosmetic formulations \cite{creton2016fracture, verdier2003effect}. 

 \begin{figure*}[t]
 \centering
\includegraphics[height=6cm]{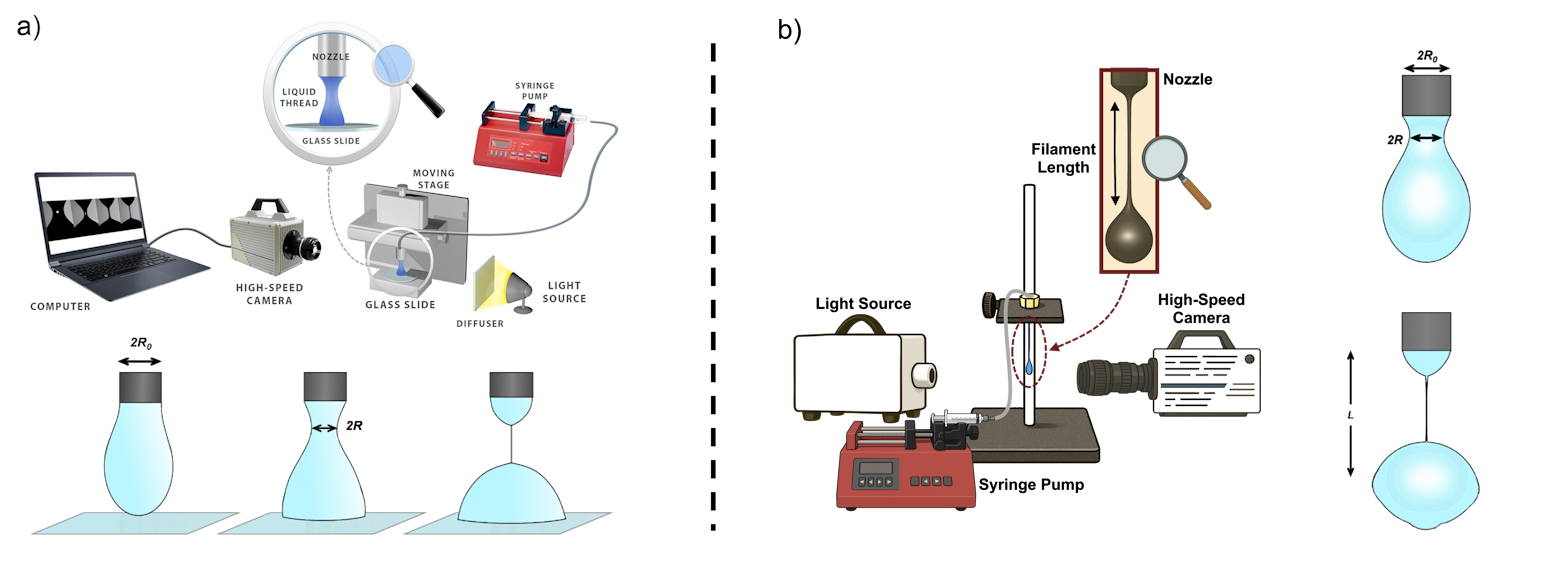}
 \caption{Two extensional rheometry techniques used to elucidate the effects of chemical structure and molecular weight on stringiness. a) Dripping-onto-Substrate (DoS) protocols capture the maximum filament lifespan and radius evolution of a neck as it thins due to capillarity and b) Dripping-into-Air (DiA) protocols allow for the visualization of a maximum filament length as well as filament radius due to gravity-driven thinning.}
 \label{fig:1}
\end{figure*}

Dripping or Dripping-into-Air (DiA) protocols can measure both filament length and lifespan, and can be used to analyze capillarity-driven pinching dynamics \cite{amarouchene_inhibition_2001, tirtaatmadja_drop_2006, rajesh2022transition}. The only study that provides detailed plots of length vs time, however, was by Cooper-White et al. \cite{cooper2002drop}, who used linear-linear axes, making the precise influence of polymers hard to discern. Like Jones and Rees \cite{jones_stringiness_1982}, later Shore and Harrison \cite{shore2005effect} in 2005 also measured the filament length for a few polymer solutions, but did not characterize an extensional rheology response. In addition to DiA, stringiness as set by extensional rheology response can be assessed by characterizing capillarity-based pinching dynamics using step-stretched capillary bridges, such as in Capillary Breakup Extensional Rheometer (CaBER) \cite{anna_elasto-capillary_2001, stelter_validation_2000, rodd_capillary_2005, calabrese2025effects}, or Trimaster \cite{vadillo_rheological_2010, vadillo_microsecond_2012}, using jetting for example in Rayleigh-Ohnesorge Jetting Extensional Rheometer \cite{sharma_rheology_2015, keshavarz_studying_2015}, and through Dripping-onto-Substrate (DoS) protocols \cite{dinic_extensional_2015,dinic_pinchoff_2017, dinic_pinch-off_2017, al_zahabi_pinching_2024}. The stretching of a liquid bridge between a finger and a thumb and similarly between two plates of CaBER, Trimaster, or their analogs create a thinner filament or neck connecting large drops at two substrates \cite{mckinley_visco-elasto-capillary_2005, vadillo_evaluation_2010, calabrese2024polymers, clasen_dispensing_2012, bazilevskii_breakup_2001, bazilevsky1990liquid, anna_elasto-capillary_2001, stelter_validation_2000}. In the most typical version, a step stain is applied, and the neck thinning or pinching dynamics are monitored after the fast retraction step is complete. The pinching dynamics of polymer solutions often displays an elastocapillary (EC) regime, with radius-time plots displaying an exponential decay. The decay constant provides a measure of $\lambda_E$, and a manifested elastic response \cite{entov_effect_1997, stelter_validation_2000, stelter_investigation_2002, anna_elasto-capillary_2001, mckinley_visco-elasto-capillary_2005, rodd_capillary_2005}. The solutions of highly flexible and extensible polymers show a terminal, viscoelastocapillary  (TVEC) regime with a linear decrease in radius. Polymers with low flexibility and extensibility sometimes display only TVEC regime. Several studies detail the protocols, limitations, challenges and successes of capillarity-based techniques \cite{sharma_rheology_2015, dinic_macromolecular_2019, dinic_rheology_2022, calabrese2025effects, zinelis2024fluid, mckinley_visco-elasto-capillary_2005}, and therefore here we focus entirely on developing an approach to classify stringiness.

In this contribution, we use the DoS rheometry protocols to characterize the extensional rheology response by visualizing and analyzing the capillarity-driven pinching dynamics of self-thinning or self-pinching necks \cite{dinic_extensional_2015, dinic_pinchoff_2017, dinic_macromolecular_2019, jimenez_extensional_2018, martinez_narvaez_dynamics_2021, kubinski_extensional_2024}. Though the DiA protocol omits the substrate under the nozzle and allows a measurement of breakup length or filament length, limitations tied to imaging resolution and recording framerates poses challenges in capturing the full filament and breakup time of extremely stringy solutions. DoS only measures filament lifespan, but due to the fixed distance between the dispensing nozzle and the substrate, videos are able to accurately capture complete pinching of the liquid neck at high frame rates. Due to the high-speed imaging challenges and the limits of optical resolution to follow changes in radius at a subpixel level, the filament lifespan ascertained with the naked eye is marginally higher than the smallest neck radius picked for analysis. Here we limit our examination of DiA protocols to polymer solutions with shear viscosity twice that of water, and rely on DoS rheometry to obtain measurement of filament lifespan, extensional viscosity and extensional relaxation time. Lastly we introduce two dimensionless groups: $V_E$ (viscoelastocapillary extensional number) and  extensional stringiness or $S_E$ that can be used to rank, and assess viscoelastic fluids. 

\section{Materials \&  Methods}

\subsection{Aqueous Polymer Solutions}

The polymers chosen for this study are commonly used as rheological modifiers in commercial and industrial applications and concentrations picked were influenced by their by chemical structure and molecular weight. Among these, Polyacrylamide (PAM, 150 kDa, BASF), Polyvinylpyrrolidone (PVP, 1300 kDa, Sigma Aldrich) and five different $M_w$ samples of Polyethylene Oxide (PEO, 300, 600, 1000, 2000, and 5000 kDa, Sigma Aldrich) represent synthetic chain-growth polymers. Two polysaccharides are included for completeness: Hydroxyethylcellulose (HEC, 720 kDa, Sigma Aldrich) and Guar Gum (GG, unknown $M_w$, BASF). HEC is a modified cellulose, whereas Guar Gum is a naturally-derived polysaccharide with complex structure. The polymers were added slowly and carefully to DI water and allowed to dissolve for over a week. During mixing, the solutions were placed on rollers for a week to avoid any chain scission that is known to occur at high deformation rates. As guar gum tends to clump and form aggregates while hydrating, its solutions were subjected to more vigorous stirring.

\subsection{Rheology and Stringiness}

Shear viscosity of the polymer solutions was characterized at 25\degree C using a concentric cylinder Couette cell geometry on an Anton Paar MCR 302 torsional rheometer. For each measurement, imposed shear rate was varied in the range of $\dot{\gamma}=$ 0.01-10\textsuperscript{3} s\textsuperscript{-1}, and the torque registered by the device was used to compute first the measured shear stress, $\tau_{12}$, and then the steady shear viscosity $\eta(\dot{\gamma}) \equiv \tau_{12} / \dot{\gamma}$. Both Dripping-onto-Substrate (DoS) and Dripping-into-Air (DiA) protocols, illustrated in Figure 1, utilize a Photron FastCam SA3 high-speed camera with a Nikon 18-55mm $f$3.5-5.6G DX lens mounted with extension tube adapters to allow for close-up imaging of small drops and thin filaments. Each solution was pumped out of a nozzle of $D=1.8288$ mm using a syringe pump set to a flowrate of $Q=0.02$ml/min. The DoS videos were captured at 19000 FPS allowing higher spatiotemporal visualization and analysis of neck pinching dynamics, while DiA videos were captured at 5000 FPS to provide an extended view of the elongated necks. The height of the video recording window is maximized to capture as much of the filament as possible. 

In dripping experiments, the filament length is measured by using time corresponding to a specified neck radius, $R/R_0 \le 0.3$. For both DoS and DiA rheometry, the high-speed videos are analyzed through MATLAB codes found in the supplementary information. For DoS, the change in filament radius and in DiA, the length of the filament are plotted against time. The radius evolution data from DoS rheometry is further analyzed to determine if the kinematics observed implies inertiocapillary (IC), viscocapillary (VC), elastocapillary (EC) or plastocapillary (PC) regime. For known surface tension $\Gamma$, the apparent extensional viscosity $\eta_E$  values are obtained by taking the ratio of capillary stress $\Gamma/R(t)$ within the pinching neck to the extension rate, $\dot{\varepsilon}$. Like the varying stress, the radius evolution data also provides a measure of $\dot{\varepsilon} = -2\dot{R}(t)/R(t)$. Lastly, extensional relaxation time, $\lambda_E$ can be calculated from the elastocapillary pinching.

\section{Results}
\subsection{Shear and Extensional Viscosity of Nine Polymer Solutions}

\begin{figure*} 
 \centering
\includegraphics[height=7cm]{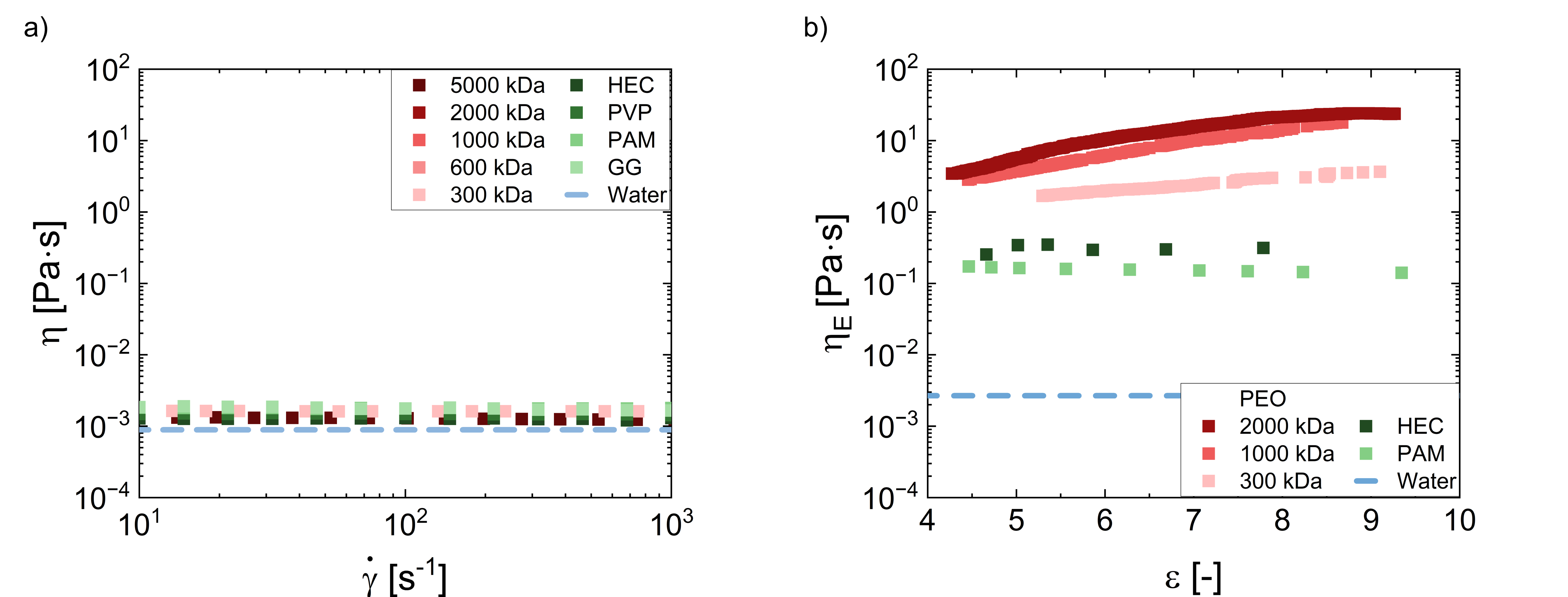}
 \caption{Shear and extensional rheology of chosen polymer solutions. a) Shear viscosity of nine polymer solutions are comparable and are roughly two times water viscosity. b) Extensional viscosity determined using DoS rheometry protocols shows significantly higher values than shear viscosity, and the overall enhancement is dependent on polymer chemistry and molecular weight.}
 \label{fig:2}
\end{figure*}

Figure 2a displays the shear viscosity as a function of shear rate for the nine aqueous polymer solutions. The shear viscosity of these solutions is rate-independent and comparable, with all $\eta_0$ nearly twice the viscosity of water. The comparable shear viscosity response is created by design as the particular concentrations were selected to be equal to or close to the overlap concentration, $c^*$, which is defined at concentration, $c$, where the volume per coil equals the equilibrium coil volume. The value of $c^*$ depends on the equilibrium coil size that, in turn, is influenced by the molecular weight, the number and the size of Kuhn segments, and polymer-solvent interactions. The absolute concentrations utilized in this study for each polymer solution are included later in Table 1.

The apparent extensional viscosity, $\eta_E$, vs Hencky strain, $\varepsilon$, shown in Figure 2b for the aqueous solutions of PEO (three $M_w$s), HEC, and PAM were determined using Dripping-onto-Substrate (DoS) protocols. Figure 2b shows the apparent extensional viscosity values determined from  $\eta_E \dot{\varepsilon}(t) = \Gamma/ R(t)$. Although the shear viscosity being more or less equal for all solutions, the extensional viscosity varies with the choice and $M_w$ of a polymer. The PEO solutions exhibit higher extensional viscosities than PVP, PAM, and HEC, with the $\eta_E$ increases with $M_w$. Furthermore, the strain-hardening response becomes much more distinct with higher molecular weight for highly flexible and extensible polymers. 

Although the shear viscosity of the five solutions was comparable, Figure 2b shows that there are demonstrably large differences in the measured values of extensional viscosity. Based on experimental and theoretical arguments first made by Trouton \cite{trouton_coefficient_1906}, higher extensional viscosity should correspond to longer lived liquid filaments and higher perceived stringiness. Since Trouton had only examined the response of Newtonian fluids, where extensional viscosity is three times shear viscosity, this difference in stringiness  of these polymer solutions with similar shear viscosity is an illustration of the influence of macromolecular concentration, molecular weight and physiochemical properties on response to extensional flow. The PAM and HEC solutions displays $\eta_E$ values which are almost 100 times or two orders of magnitude larger than their respective shear viscosities, $\eta$. For PEO 2000 kDa, the Trouton rato exceeds $Tr=\eta_E/\eta>10^4$. To assess how difference in shear and extensional viscosity influences apparent stringiness, we carried out falling drop or Dripping-into-Air experiments, discussed next. 

\subsection{Filament Length via Dripping-into-Air (DiA) Protocols}

\begin{figure*}[t]
\centering
\includegraphics[height=14cm]{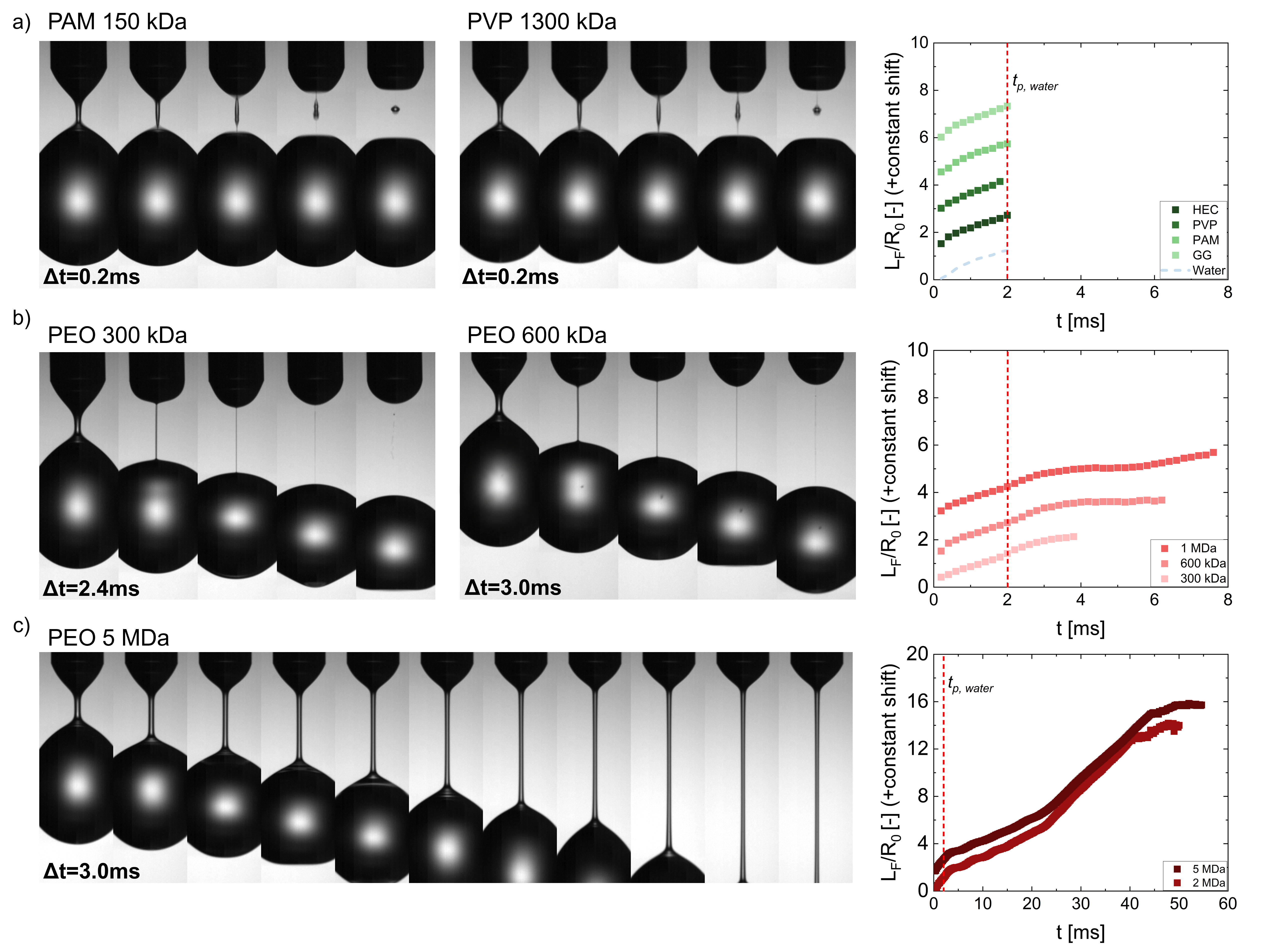}
 \caption {Dripping-into-Air (DiA) experiments for polymer solutions with comparable shear viscosity allow the visualization of and analysis of filament length. The length vs time plots are included in the last column. Image sequences for aqueous solutions of  (a) PAM ($M_w$ = 150 kDa) and PVP ($M_w$ = 1300 kDa) respectively show rapid pinching, and satellite droplet formation; b) PEO with $M_w$ = 300 KDa and 600 kDa respectively show delayed pinching and formation of long, thin, persistent filaments. and  c) PEO with $M_w$ = 5000 KDa show significantly longer maximum filament length and filament lifespan. The PEO solutions are significantly more stringy and increasing their molecular weight results in longer filament length and filament lifespan.  Furthermore, the image sequences for PEO solutions reveal that the at matched filament length and after matched timespan, solutions with higher molecular weight show thicker filaments.}
 \label{fig:3}
\end{figure*}

Image sequences obtained by Dripping-into-Air displayed in Figure 3 display contrast in the evolution of neck shape and filament length. Images are included for nine polymer solutions: five different molecular weights of PEO and one solution each of HEC, PVP, PAM and Guar Gum, all with comparable shear viscosity. For each sequence, the first snapshot is displayed for matched value of $R/R_0$. The plots included in the last column display filament length, $L_F$ normalized by the nozzle radius, $R_0$ as a function of time, $t$. To identify the filamentous part of the pendant drop on track to pinch-off, we chose to define necked region with $R/R_0 < 0.3$ as the filament. The plots, shifted vertically by a constant value to avoid overlap, show that the growth of the filament depends on macromolecular properties. In the first row with images and plots labeled Figure 3a, solutions of PAM and PVP illustrate rapid pinching of the liquid neck that displays tell-tale signs of inertiocapillary (IC) breakup with a sharp cone at the end of the filament. The short and short-lived filaments detach from the falling drop at the tip of the cone formed at their lower end. A single visible satellite drop appears after the pinch-off is completed. The last column shows the filament length evolution and for solutions of PAM 150 kDa and PVP 1300 kDa the filament length is comparable to water, the solvent, despite the solution viscosity being nearly twice as large. On comparing the images and dynamics for PVP and PAM in Figure 3a to those in Figure 3b shown for PEO 300 and 600 kDa solutions of comparable viscosity, the dramatic change in neck shape, lifetime, and length are revealed. 

The image sequences in Figure 3b for 300 kDa and 600 kDa are shown with a lower different spatiotemporal resolution to emphasize the persistent nature of the filament. Both PEO solutions form slender cylindrical necks, similar to the neck shapes realized in the DoS experiments and characteristic of a visible elastocapillary response. Increasing the molecular weight of the PEO solutions further increases filament lifespan, $t_f$ and final filament length, $L_{F}$ as evident from Figure 3c. If all the drops were dispensed from the nozzle at the same instant, the image sequences show that after similar timespan or number of snapshots, the highest $M_w$ PEO displays the thickest and longest filaments, that appear to thin at a slower rate. The plot displaying filament length evolution of the aqueous solution of PEO 2000 kDa and 5000 kDa requires a substantially expanded axis range. The 5000 kDa solution formed filaments that persisted and stretched beyond the recording window captured in the Dripping-into-Air (DiA) experiments. In addition to filament length and lifespan, the rate of pinching of fluid necks also influences the perception of stringiness, and the filament radius for the two higher $M_w$ PEO solutions is observed to be larger after equal elaspsed time, and decrease at a slower rate than the 300 and 600 kDa PEO image sequences.

The visualization of the shape evolution of the highly extended filaments formed for PEO 2000 kDa and 5000 kDa illustrates the challenges inherent to the DiA of highly viscoelastic fluids. The first hurdle stems from a limited camera visualization window, or the actual recordable width and length of a video that need to be captured at a sufficiently high frame-rate and spatial resolution. The longer the maximum filament length becomes, the more difficult it is to keep the full length of the filament in frame while still capturing filament evolution at a high spatiotemporal resolution for the analysis of the pinching neck. There appears to be an inherent law of conservation of misery which limits the spatiotemporal resolution achievable for highly elastic and highly viscous fluids. Long-lived threads also form for high viscosity fluids like honey, and indeed such threads are susceptible not only to the Rayleigh-Plateau instability, but also to the influence of ambient conditions (temperature variations, air drafts, etc.) and the problem of drying. Furthermore, if the camera lens is zoomed out in an attempt to capture the full filament length, the amount of representative pixels capturing the filament in video is reduced, greatly impacting the sensitivity of the video processing code. Despite these hurdles, DiA remains a viable option for qualitatively observing and quantitatively analyzing the manifest stringiness of polymer solutions.

\subsection{Contrasting Pinching Dynamics using DoS Rheometry}

Image sequences included in Figure 4a illustrate the variation in the shapes and radius of necks undergoing capillarity-driven pinching in stretched bridges created and visualized by dripping-onto-substrate (DoS) protocols. The images acquired using high-speed imaging are shown with a matched time step of 1.5 ms. A longer sequence of frames implies a longer filament lifespan. Initially, an hourglass shaped liquid bridge is formed between the nozzle and a partially-wetting (or pinned) substrate, as visible in the first image of each row in Figure 4a. Close inspection of the neck shapes shows that the liquid bridges of HEC, GG, PAM and PVP solutions formed slightly rounded conical necks that tapered to a small nib-like point similar to that of ink pens. These neck shapes are quite similar to the conical necks formed by low viscosity Newtonian fluids, including water that is used as a solvent for preparing the nine polymer solutions. However, several of the polymer solutions depart from a standard conical neck, producing a very short and thin filament between the cone and the sessile drop. The filament appears slightly longer and thicker in diameter in PVP and HEC solutions. The neck shapes for PEO solutions including for $M_w$ = 300 kDa and 600 kDa display the formation and evolution of a well-defined slender cylindrical neck. The filament lifespan increases with the molecular weight of the PEO.

\begin{figure*}[t]
 \centering
 \includegraphics[height=14cm]{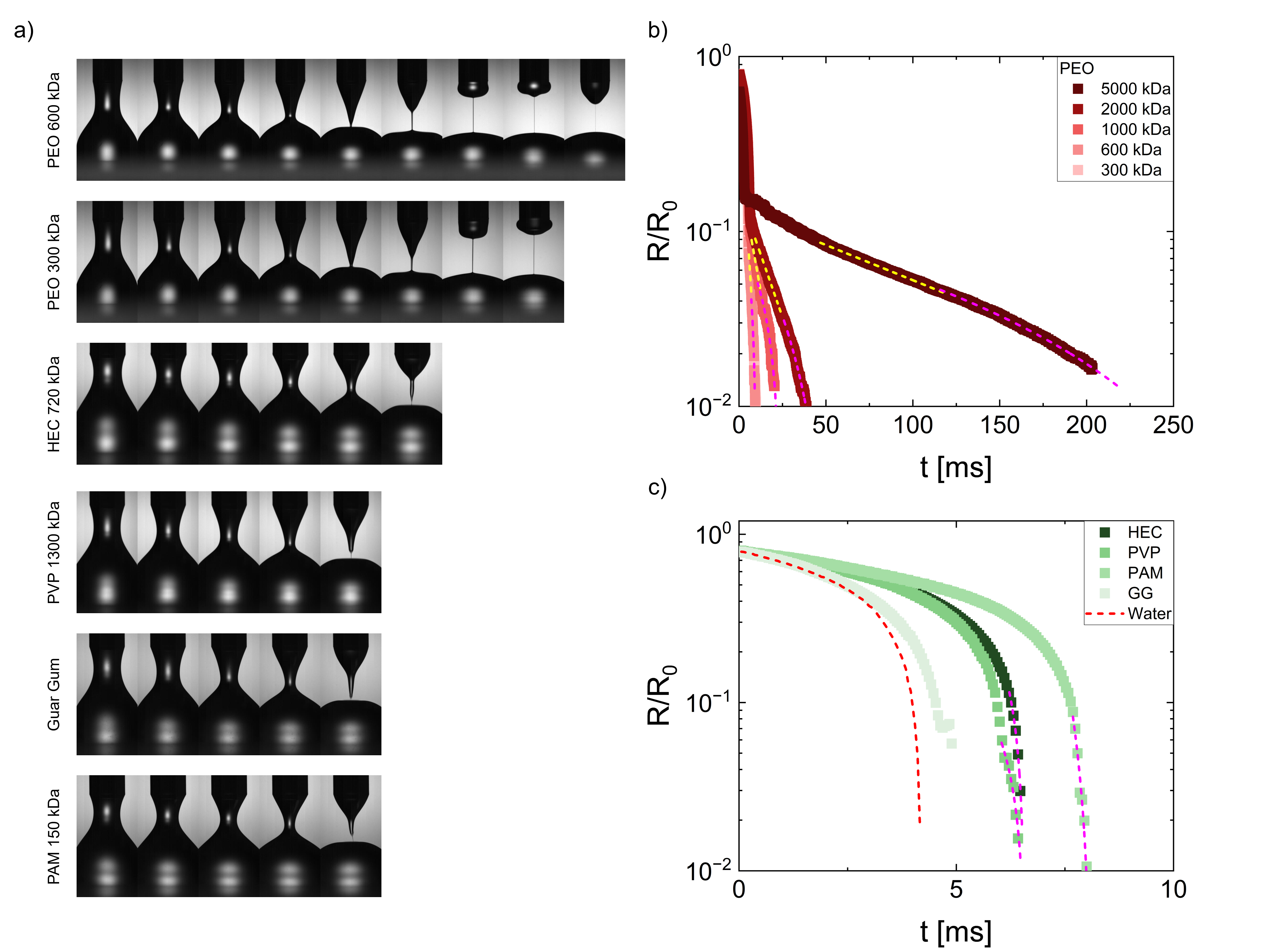}
 \caption{Dripping-onto-Substrate (DoS) experiments for polymer solutions with comparable shear viscosity showing the shape and radius evolution of the pinching neck. a) Image sequences show the shape evolution of liquid bridges formed between a fixed nozzle and a substrate, with necks undergoing capillarity-driven pinching. The polymer solutions show distinct neck shapes and filament lifespans. Snapshots shown for each set are $\Delta t = 1.5ms$ apart. b) Radius evolution over time for the five comparable shear viscosity solutions of PEO with five distinct molecular weights, shows that the EC regime and filament lifespan increase with $M_w$. c) Evolution of scaled radius with time shows that the aqueous solutions of PAM, GG, HEC, and PVP pinch-off with relatively short filament lifespan, and the EC regime is either not manifested or quite short-lived.}
 \label{fig:4}
\end{figure*}

The radius evolution data, extracted from the analysis of image sequences for water and nine polymer solutions, are included in Figures 4b and 4c. Radius vs. time data for water, which is a relatively low viscosity, Newtonian fluid, exhibits an inertiocapillary (IC) response, described by the following expression:

\begin{equation}
\begin{aligned}
  \frac{R(t)}{R_0} = \alpha \space (\frac{\Gamma}{\rho R_0^3})^{1/3} \space (t_p-t)^{2/3}= \alpha \space (\frac{ t_b}{t_{ic}})^{2/3}
\end{aligned}   
\end{equation}

The expression includes $\alpha$, an O(1) numerical coefficient and a pinch-off time, $t_p$. The expression also includes shows that the radius evolution is driven by an interplay of capillary and inertial stresses, that set the characteristic timescale for the pinching often described as inertiocapillary time, $t_{ic}=t_{{\Gamma}{\rho}}=({\Gamma}/{\rho R_0^3})^{1/2}$. This $t_{ic}$ is same as the Rayleigh timescale which obtained first by dimensional arguments in Rayleigh's analysis of jetting of inviscid fluid. Thus the scaled radius evolution can be described in terms of a scaled backwards time, $t_b={t_p-t_c}$.

Figure 4b and 4c show IC fits for water and all polymer solutions. For aqueous PEO solutions shown in Figure 4c, the initial IC response transitions to a distinct second regime, due to the influence of viscoelastic stresses generated by the response of stretched polymers within the pinching neck. The interplay of capillary and viscoelastic stresses leads to elastocapillary (EC) response, associated with an exponential decay in radius, that appears linear on a semi-log plot. Denoting the transition point at $R_c$ and $t_c$, the radius evolution in the EC regime can be captured by the following expression that includes the extensional relaxation time, $\lambda_E$ as a decay constant:

\begin{equation}
  \frac{R(t)}{R_0} = \frac{R_c}{R_0} \: \text{exp} (-\frac{t-t_c}{3 \lambda_E})
\end{equation}

The earlier version of the expression for elastocapillary pinching, derived by Entov and Hinch in 1997 using the Oldroyd-B model used $t_c=0$ and determined $R_c= R_0({GR_0}/{\Gamma})^{1/3}$ which was derived assuming no tension in the filament \cite{entov_effect_1997}. Subsequently, Clasen et al in 2006, and many others, including Eggers, found $t_c=0$ and said $R_c= R_0({GR_0}/{2\Gamma})^{1/3}$ on including the tension term \cite{clasen_how_2006, eggers_nonlinear_1997, clasen_beads--string_2006, gaillard2025elastic}. Dinic and Sharma favor Eq. 2 as it explicitly includes only one fit parameter, which is equated to extensional relaxation time. Dinic and Sharma argued that the IC-EC transition is sharp for dilute solutions of highly extensible, highly flexible polymers like PEO, but the lower extensibility and flexibility polymers or nondilute solutions of all polymers display a softer or less sharp transition. Figure 3b includes the dataset for HEC solution, inspired by the Dinic and Sharma study \cite{dinic_flexibility_2020}, and here the EC regime is shown with a dotted line. The PVP solution also shows a similar EC response, but the EC regime is not observed for the corresponding solutions of the relatively low $M_w$ highly flexible PAM and the semi-flexible polysaccharide guar gum. 

For all the radius evolution plots where an EC regime manifested, $\lambda_E$ values were extracted. The HEC solutions exhibited the shortest $\lambda_E$, followed by the PVP solution, and finally the PEO solutions with increasing $M_w$ of PEO in succession. The $\lambda_E$ values for Guar Gum and PAM could not be calculated from the data. As all solutions were prepared in the same solvent, and polymer contributions to the solution viscosity were comparable, here, the higher value of $\lambda_E$ leads to longer filament lifespan.

The PEO solutions as well as HEC and PVP solutions display a linear decrease in radius before pinching due to the interplay of capillary stresses with viscoelastic stresses contributed by highly stretched polymers. In the FENE-P model, the viscoelastic stresses are said to saturate due to the finite extensibility of highly stretched macromolecules. A terminal, steady extensional viscosity, $\eta_E^\infty$, that is independent of strain $\epsilon_H$ and strain rate can be obtained from the analysis of this linear regime using the following expression for the terminal, viscoelastocapillary (TVEC) regime:

\begin{equation}
  \frac{R(t)}{R_0} = {X_{vc}} (\frac{t_f-t}{t_{tvec}})={X_{vc}} (\frac{t_p}{t_{tvec}})
\end{equation}

A linear decrease in radius is also observed in relatively high viscosity Newtonian fluids due to the interplay of viscous and capillary stresses. For Newtonian fluids, by defining a  viscocapillary time $t_{vc}=\eta_0 R_0/\Gamma$ replaces $t_{tvec}$. Historically, the prefactor, $X_{vc}$ for the Newtonian fluids was given out to be ~1/6 by the Papageorgiou's theory \cite{papageorgiou_breakup_1995}, and was found to be dictated by the experimental parameters as detailed by McKinley and Tripathi \cite{mckinley_how_2000}. The ratio of viscocapillary (VC) to inertiocapillary (IC) timescales is a dimensionless measure of viscosity called Ohnesorge number, $Oh=\eta_0/\sqrt{\rho\Gamma R_0)}$. Only VC thinning is manifested if Oh > 1 and thus if $R_0$<$l_{Oh}$ where $l_{Oh}=\eta_0^2/{\rho\Gamma}$. However, for low viscosity fluids, typically IC regime is observed but IC transitions to the viscous regime must arise before breakup, though it is manifested only if $l_{Oh}>l_{res}$ where the optical resolution is set by optics used. Thus, three different timescales govern the pinching dynamics of unentangled polymer solutions: the inertiocapillary or Rayleigh, timescale, $t_{ic}=((\rho R_0^3)/\Gamma)^{1/2}$, the viscocapillary timescale, $t_{vc}=(\eta R_0)/\Gamma$, and the extensional relaxation time, which can be obtained using DoS and other capillarity-based pinching studies. The radius evolution for these polymer solutions often displays three regimes, including an intermediate regime with exponential delay (the EC stage). The early stage is either IC or VC and this can be known a-priori by computing the dimensionless Ohnesorge number \textit{(Oh)}, which indicates whether $t_{vc}$ or $t_{ic}$ is dominant. The elastic effects are often estimated by computing the Zimm relaxation time, which represents the diffusion time of an unperturbed single coil. For dilute polymer solutions, the variation in shear relaxation time is usually not measurable, whereas the extensional relaxation time can show the influence of solvent quality and macromolecular properties especially extensibility. A wealth of experimental studies carried out using DoS rheometry, with supplementary datasets from CaBER and dripping, claim that the extensional rheological response captured by $\lambda_E$ and $\eta_E^{\infty}$ are correlated with sprayability, spinnability, jettability and processing behavior. The connection requires a careful assessment of the stretched chain properties. For example, the linear stability analysis of viscoelastic jets shows that at a matched viscosity, the presence of elasticity makes a jet more susceptible to breakup by small perturbations \cite{middleman1965stability, kroesser1969viscoelastic, ardekani_dynamics_2010, sharma_rheology_2015}, suggesting that the breakup length could be smaller. However, in the late regime, as a neck thins down considerably, the assumptions made for the linear stability analysis do not apply, and the nonlinear flow regime is extensively determined by extensional rheology response \cite{christanti_surface_2001, christanti_effect_2002, clasen_beads--string_2006, sharma_rheology_2015}.

\section{Discussion}
\subsection{Filament lifespan and macromolecular properties}

\begin{table*}[t]
\caption{Representative values of zero-shear viscosity, terminal steady extensional viscosity, extensional relaxation time, and filament lifespan obtained by shear and extensional rheology characterization of polymer solutions.}
\label{tab:1}
\squeezetable
\begin{ruledtabular}
\begin{tabular}{cccccccccc}
Polymer & $M_w$ (kDa) & $c^*$ (wt\%) & $c/c^*$ & $\eta_0$ (Pa s) & $\eta_E^\infty$ & $\lambda_E$ (s) & $t_f$ (s) & $l_{Oh}$ ($\mu$m) & Ref. \\
\midrule
PEO & 600 & 0.25 & 1 & 0.00164 & 3.37 & 0.94 & 10 & 0.037 & This Study \\
PEO & 2000 & 0.1 & 1 & 0.00161 & 5.02 & 4.98 & 39 & 0.036 & This Study \\
PEO & 5000 & 0.048 & 1 & 0.00129 & 27.38 & 12.26 & 203 & 0.023 & This Study \\
PEO & 1000 & 0.2 & 1.18 & 0.0019 & 21 & 2.80 & 30 & 0.087 & Dinic \cite{dinic_flexibility_2020} \\
HEC & 720 & 0.17 & 1 & 0.0022 & 0.40 & 0.22 & 4 & 0.067 & Dinic \cite{dinic_flexibility_2020} \\
NaCMC & 250 & 0.03 & 2 & 0.055 & 1.20 & 0.70 & 13 & 41 & Jimenez \cite{jimenez_capillary_2020} \\
PAM & 1000 & 0.32 & 0.26 & 0.072 & 7.1 & 3.1 & 25 & 70 & Soetrisno \cite{soetrisno_concentration_2023} \\
\end{tabular}
\end{ruledtabular}
\end{table*}

Filament lifespan, $t_f$ obtained from DoS and other capillarity-driven pinching studies, provides a metric correlated to the observation of a longer lived liquid bridge, often associated with a larger stringiness or stickiness. The value of $t_f$ presents no information about the pinching kinematics or dynamics (IC, VC, EC or PC), or the underlying macromolecular properties, but in one glance, a comparison of these values provides an evaluation of the apparent stringiness. For Newtonian fluids, increasing shear viscosity leads to slower capillarity-driven pinching, and higher values of $t_f$ in DoS rheometry (and CaBER studies), as well as increasing final filament length, $L_{F}$, before pinch-off. The $t_f$ value can be associated with the time needed to reach the minimum radius resolvable by available optics, and $t=0$ can be defined as an early stage of thinning - in this study $R/R_0=0.7$. Alternatively, $t_f$ can be obtained from the IC, VC or PC fit that captures the last stage of breakup. In Figure 5a, filament lifespan, $t_f$ is plotted against zero shear viscosity, $\eta_0$. For the polymer solutions  characterized in Figures 2-4, the highlighted region of the plot in Figure 5a shows that $t_f$ values can be quite distinct, despite similar $\eta_0$. Increasing the $M_w$ of PEO solutions but keeping $\eta_0$  matched by an appropriate concentration reduction, leads to an increased filament lifespan. 

In order to decipher how $t_f$ correlates with macromolecular properties and measured rheological response, we compile representative datasets from published DoS rheometry studies \cite{jimenez_capillary_2020, jimenez_extensional_2018, dinic_flexibility_2020, soetrisno_concentration_2023}. The HEC 720 kDa and PEO 1000 kDa data is drawn from a DoS rheometry investigation by Dinic and Sharma \cite{dinic_flexibility_2020} of two series of polymer solutions with similar coil size, and comparable overlap concentration, implying a similar Zimm time in the single coil limit. As expected, the dilute PEO and HEC solutions showed comparable shear viscosity at matched concentration. However, radius evolution data obtained using DoS rheometry shows a dramatic difference once elastocapillary regimes manifest. The elastocapillary timespan, filament lifespan, extensional relaxation time and extensional viscosity are much larger for the PEO solution than HEC solution due to the significantly higher extensibility \cite{dinic_flexibility_2020}. Two datasets are for solutions of polyelectrolytes \cite{jimenez_capillary_2020, jimenez_extensional_2018}: celullose gum of sodium carboxymethylcellulose (NaCMC), a charged polysaccharide commonly used as food thickener, and poly(acrylic acid) or PAA, example of a rather well-studied model polyelectrolyte. The dataset for poly(acrylamide) or PAM was acquired in solutions with glycerol-water mixture as a solvent. The aqueous PVP solution data is extracted from another DoS rheometry investigation. The persistence of the filament increases as the data points shift to the upper right corner of the plot. Increasing polymer concentration typically increases the filament lifespan, $t_f$, although the relative rise is dictated by macromolecular choice and properties (molecular weight, chain chemistry, polymer-solvent interaction, electrostatic and hydrodynamic interactions, etc). For a Newtonian fluid, the knowledge of shear viscosity is adequate, and raising shear viscosity makes filaments live longer. However, Figure 5a shows that shear viscosity alone is not a robust metric for judging filament lifespan.

\begin{figure*}[t]
\centering
\includegraphics[height=14cm]{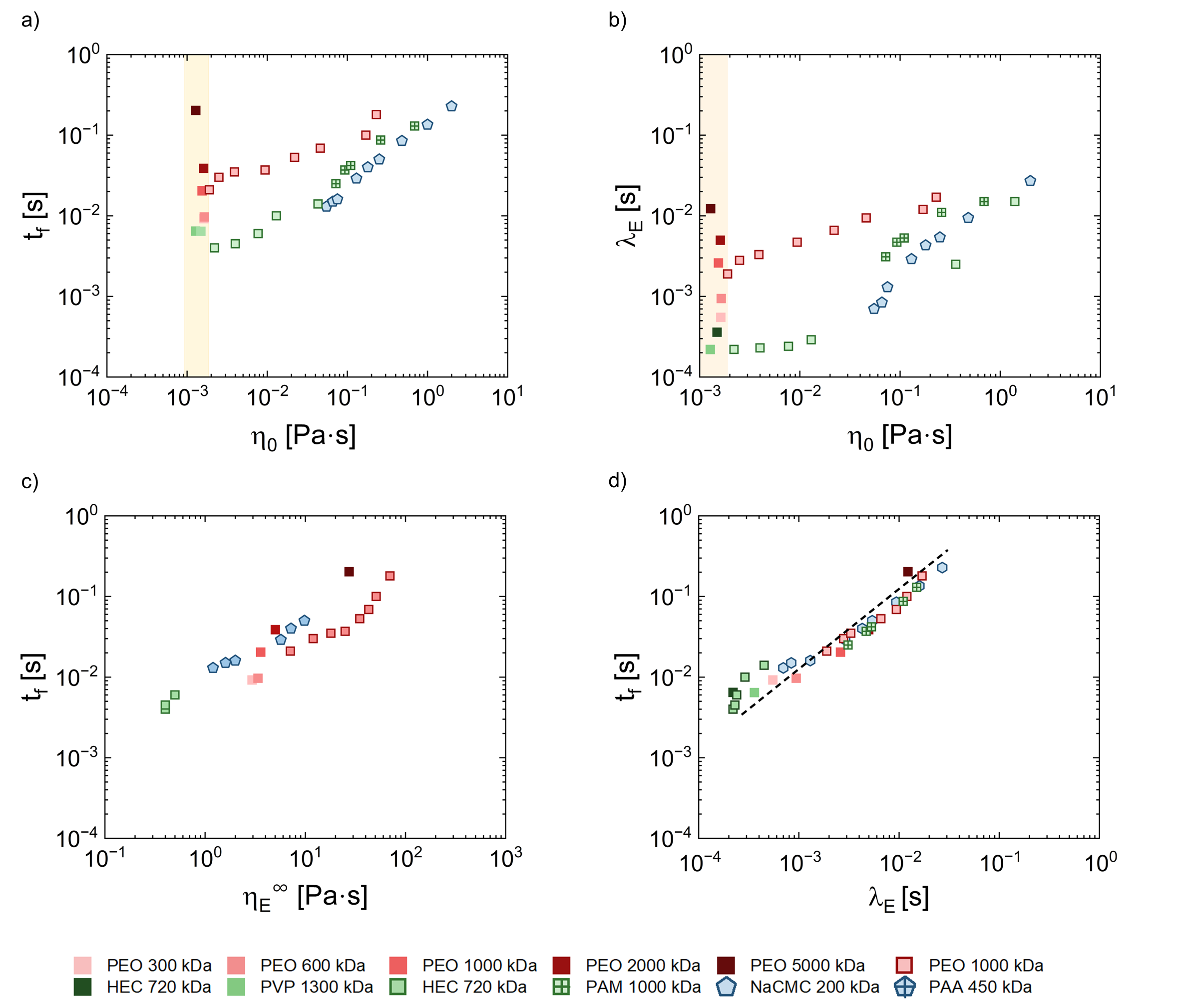} \caption{Correlation between filament lifespan, $t_f$, and the shear and extensional rheology response of polymer solutions. a) Filament lifespan plotted against $\eta_0$, shows that polymer solutions of comparable zero shear viscosity can display widely different rates and timescales for pinching. In addition to $t_f$ measured in this study (banded region), the plot includes datasets extracted from several published studies including Dinic, et al (2019), Jimenez, et al (2018, 2020), and Soetrisno, et al (2023). b) The extensional relaxation time, $\lambda_E$ can vary considerably even for shear viscosity matched polymer solutions, implying like  $t_f$, $\lambda_E$ is not correlated with $\eta_0$. c) The persistence of the filament, captured by filament lifespan, increases with steady, terminal extensional viscosity. d) The $t_f$ vs $\lambda_E$ plot shows that all the datasets exhibit a strong correlation. Higher $\lambda_E$ invariably leads to longer elastocapillary regime, and the corresponding dependence on macromolecular properties can be used to control both $t_f$ and $\lambda_E$}
\label{fig:5}
\end{figure*}

Figure 5b plots extensional relaxation time $\lambda_E$ against zero shear viscosity, $\eta_0$. Again, while the shear viscosities were matched for the highlighted solutions, $\lambda_E$ was short for the solutions which exhibited Newtonian-like behavior in Figures 2-4, but increased with increasing $M_w$ for the PEO solutions. Incorporating datasets from other studies resulted in a plot which showed little correlation between $\lambda_E$ and $\eta_0$; however, the datapoints created a pattern similar to that seen in Figure 5a, which suggests a strong relationship between $t_f$ and $\lambda_E$. Furthermore, Figure 5c plots filament lifespan, $t_f$ against the steady, terminal extensional viscosity, $\eta_E^{\infty}$ for the viscoelastic solutions showing TVEC pinching with linear decrease in radius over time. For solutions made with same polymer, the $t_f$ shows a concentration-dependent increase in $\eta_E^{\infty}$ for fixed $M_w$ and for fixed concentration, a $M_w$ dependent increase. The polysaccharides appear to have comparable filament lifespans to PEOs of higher molecular weight. Figure 5d shows the plot of filament lifespan, $t_f$ against the extensional relaxation time, $\lambda_E$ for the viscoelastic solutions that show elastocapillary pinching. Here the $t_f$ shows a nearly linear correlation to the $\lambda_E$ and in contrast with the  $t_f$ vs $\eta_0$, Figure 5d shows a rather impressive collapse. Zero shear viscosity $\eta_0$ quantifies resistance to flow in the limit of low shear rate, or low or no deformation, representative of response to weak shear flows which weakly perturb polymer coils. The dilute solution rheology is primarily set by the number and the size of polymer coils. In contrast, $\lambda_E$ is representative of stretch polymer hydrodynamics, for polymer chains can undergo substantial stretching and orientation in response to strong extensional flow fields.

Dinic and Sharma argued that the ease and extent of stretching, dynamics of stretched chains, and the time needed to relax back on cessation of extensional flow depend on three macromolecular properties: flexibility, extensibility and segmental dissymmetry, all ultimately dictated by polymer chemical structure, molecular weight and polymer-solvent interactions \cite{dinic_flexibility_2020}. Flexibility reflects how readily a polymer chain undergoes conformational deformation, and is influenced by complex chemical structure as well as any inter/intramolecular interactions. The extensibility parameter, $L_E$ describes the magnitude by which a chain can stretch, calculated as the ratio of fully stretched chain length (or contour length), to the length of the unstretched chain. As chains become longer and as $L_E$ increases, their capacity to store elastic stresses and ultimately, $\lambda_E$ increases as well. Thus, in the comparison of Figure 5, $\lambda_E$ provides a more meaningful correlation to filament lifespan and stringiness versus shear viscosity. Figure 5d illustrates that the filament lifespan is nearly one order of magnitude larger than the extensional relaxation time. Table 1 provides a summary of the parameters inferred from shear and extensional rheology characterization of the polymer solutions, and includes  representative examples from previously published studies. Both $\lambda_E$ and $t_f$ increase with increasing $M_w$ of PEO. Dinic and Sharma showed that the solutions of HEC (720 kDa) and PEO (1000 kDa) have similar overlap concentration and $\eta_0$ for dilute solutions at matched polymer concentration, yet the PEO solutions show distinct extensional relaxation time that of is an order of magnitude higher than that of HEC, while the extensional viscosity is two orders of magnitude larger for PEO.

\subsection{Viscoelastocapillary Extensional $Oh-De$ or $V_EDeOh$ Stringiness Mapping in a Pinch}

\begin{figure*}[t]
\centering
\includegraphics[height=14cm]{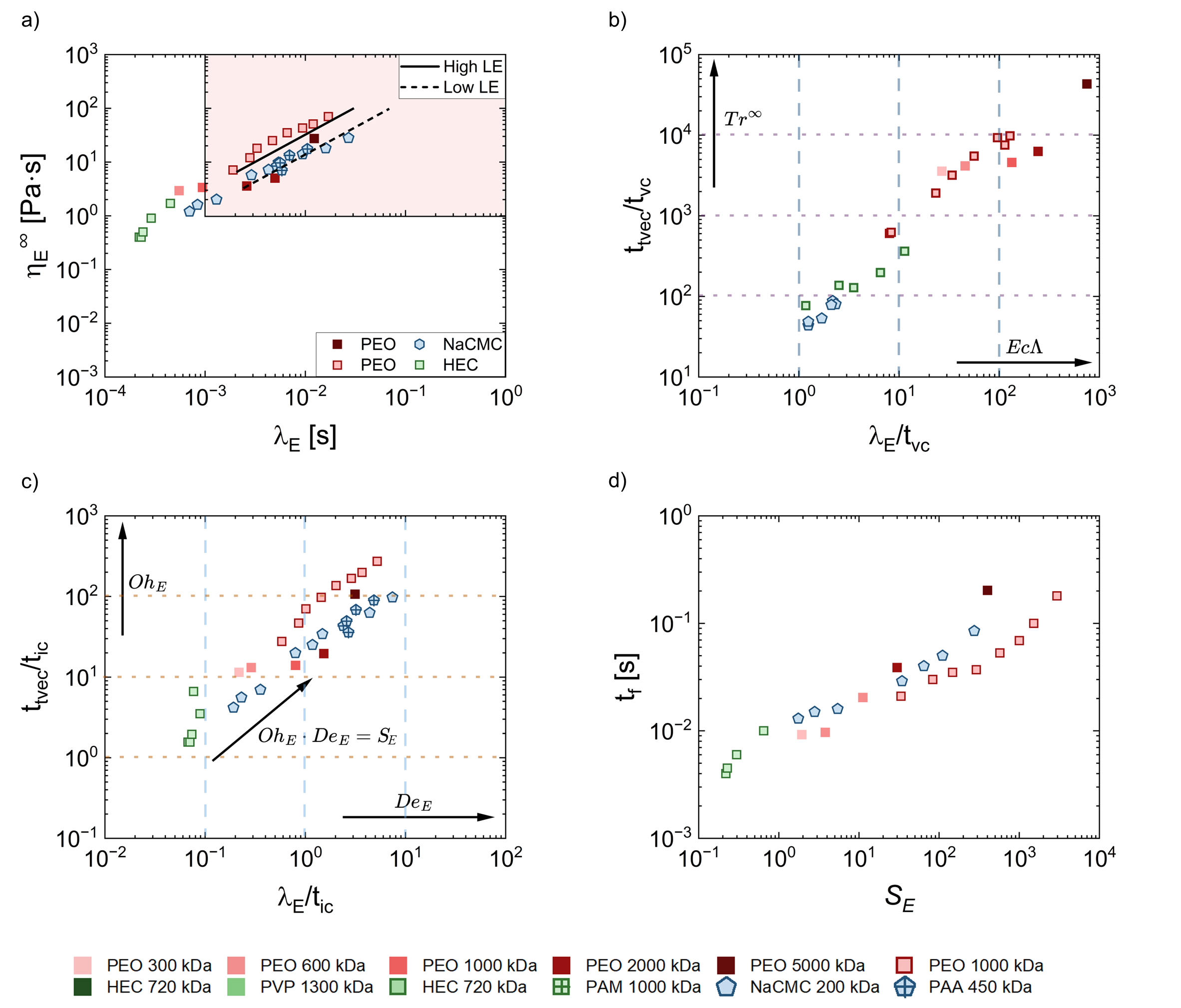} \caption{Rheologically mapping extensional Oh-De stringiness of polymer solutions. a) The $\eta_E^{\infty}-\lambda_E$ plot, that includes shows the original Stelter or the Stelter-Brenn-Yarin-Singh-Durst version as the shaded region, plots $\eta_E^\infty$ against $\lambda_E$. The steady, terminal extensional viscosity shows a better correlation with $\lambda_E$ and correspondingly with filament lifespan. The plot illustrates influence of polymer structure and molecular weight on stringiness, with two limiting curves, identified by Stelter et. al for flexible coils (solid line) and rigid rods (dashed line). Datapoints for HEC and both NaCMC and PAA in neutral solvent trend closer to rigid rod behavior, while PEO solutions cluster closer to flexible chain behavior. b) The plot compares two scaled timescales $t_{tvec}/t_{vc}$ and $\lambda_E / t_{vc}$. Horizontal dotted lines represent increases in $Tr^\infty$, while vertical dotted lines represent increases in $Ec_\Lambda$. The product of the ordinate and abscissa parameters can be understood as an extensional viscoeolastocapillary number or $Ve$.  c) Multiplying $Oh$ by the Trouton Ratio $Tr^\infty$ demonstrates an effective extensional viscosity measure which, when plotted against $De_E$, collapses datasets onto a curve with a slope of 1. Dashed lines are drawn between the axes to denote Stringiness Factor, $S_E$, with increases in $Oh_E$ and $De_E$ inherently tied to increasing stringiness factor. All experimental data obtained for this study is highlighted in yellow. d) The $t_f$-$S_E$ plot illustrates that the filament lifespan increases with stringiness factor of a polymer solution.}
\label{fig:6}
\end{figure*}

\begin{table*}[t]
\caption{Relevant timescales and dimensionless numbers used in mapping stringiness.}
\label{tab:2}
\squeezetable
\begin{ruledtabular}
\begin{tabular}{ccccccccccc}
Polymer & $M_w$ (kDa) & $t_{ic}$ (ms) & $t_{vc}$ (ms) & $t_{tvec}$ (ms) & $Tr^\infty$ & $Oh$ & $Oh_E$ & $De_E$ & $L_E^2$ & $S_E$ \\
\midrule
PEO & 600 & 3.26 & 0.021 & 85.60 & 2054.88 & 0.20 & 13.13 & 0.28 & 2350.33 & 3.78 \\
PEO & 2000 & 3.26 & 0.020 & 127.51 & 3118.01 & 0.20 & 19.56 & 1.53 & 6945.77 & 29.90 \\
PEO & 5000 & 3.26 & 0.016 & 695.45 & 21224.81 & 0.16 & 106.71 & 3.76 & 15844 & 401.47 \\
PEO & 1000 & 1.89 & 0.022 & 211.67 & 4800.00 & 0.37 & 56.12 & 1.48 & 3720 & 83.33 \\
HEC & 720 & 1.89 & 0.019 & 70.56 & 181.82 & 0.33 & 1.87 & 0.12 & 46 & 0.22 \\
NaCMC & 250 & 2.22 & 0.56 & 24.6 & 21.82 & 7.67 & 5.54 & 0.32 & -- & 1.75 \\
PAM & 1000 & 2.22 & 0.73 & 260.73 & 176.79 & 10.04 & 58.81 & 1.40 & 837--1255 & 82.23 \\
\end{tabular}
\end{ruledtabular}
\end{table*}

Observations of higher stringiness are associated with longer filament lifespan, $t_f$ in DoS rheometry, and longer final filament length, $L_{Ff}$ as well as $t_f$ in dripping-into-air (DiA) experiments. Next, we examine material properties and rheometric measures to construct a map of the  more stringy solutions than less stringy ones. In Figure 6a, shear viscosity, $\eta_0$ is plotted against extensional relaxation time, $\lambda_E$ for polymer solutions characterized in this study, supplemented by the same studies as used the plotting filament lifespan, $t_f$ against viscosity in the previous section. The juxtaposition of plots of $t_f-\eta_0$, $t_f-\lambda_E$,  $\eta_0-\lambda_E$ and $\eta_E^{\infty}-\lambda_E$ suggests that the highest $t_f$ are observed for polymer solutions with the longest $\lambda_E$ and highest $\eta_E^{\infty}$ values.

Figure 6a retains the extensional relaxation time, $\lambda_E$  from the EC regime on the x-axis as a measure of $t_f$ and material properties. It includes steady, terminal extensional viscosity, $\eta_E^{\infty}$ from the TVEC regime on the y-axis. The plot includes a shaded pink region and two parallel lines that pay homage to the $\eta_E^{\infty}$-$\lambda_E$ plot first described by Stelter, et al., in 2001, with measurements obtained using a snap-stretched liquid bridge (CaBER-style device). According to Stelter et al, the two lines captured the behavior of solutions of flexible and stiff polymer chains \cite{stelter_investigation_2002}. In this figure we reframe the two lines in terms of distinct extensibility, $L_E$. Polymer solutions that lie on the low $L_E$ curve include polyelectrolytes such as NaCMC and PAA while datapoints for linear, high $L_E$ PEO fall along the higher curve. Although the solutions tested in the original dataset were limited by the inherent limitations of step-stretch tests like CaBER \cite{rodd_capillary_2005}, which cannot measure low viscosity solutions ($\eta_0$ < 20 mPa$\cdot$s) with short ($\lambda_E$ < 1ms), the inclusion of DoS data extrapolates this framework beyond the ranges measured by Stelter, et. al. and further illustrates the differences between low $L_E$ and high $L_E$ polymer chains \cite{stelter_investigation_2002}.

 We seek to cast $\eta_E^{\infty}-\lambda_E$ in a form that represents the departure from the inertiocapillary (IC) or viscocapillary (VC) pinching in the clearest fashion, and provide a good measure of stringiness. Therefore, first we recast the term $\eta_E^{\infty}$ as the terminal visco-elastocapillary timescale, $t_{tvec} = 2\eta_E^\infty R_0 / \Gamma$, which describes late-stage thinning of a viscoelastic fluid as polymer chains approach finite extensibility and create two dimensionless groups by scaling both $t_{tvec}$ and $\lambda_E$ first with the viscocapillary ($t_{vc}$) timescale, $t_{vc} = \eta_0 R_0/ \Gamma$, which describes pinching of a viscous Newtonian fluid. On the y-axis, the ratio $t_{tvec}/t_{vc}=\eta_E^{\infty}/\eta_0=Tr^{\infty}$ captures the contrast between the steady, terminal extensional response from highly extended chains and the steady shear response of the near equilibrium coils. In the limit of ultralow concentration, the extensibility parameter, $L_E$ computed using single chain statistics determines the value of $Tr^{\infty}$. The x-axis involves a ratio of $\lambda_E/t_{vc}$ which can be interpreted as a product of two terms: $Ec \Lambda$, representing the dimensionless group known as elastocapillary number, $Ec=\lambda \Gamma/\eta_0 R_0$ and the ratio of extensional to shear relaxation times, $\Lambda=\lambda_E/\lambda$. The product of abscissa and ordinate parameters can be thought of a dimensionless group that we call Viscoelastocapillary Extensional number or $V_E=Tr^{\infty}_EEc\Lambda=\lambda_E\eta_E^{\infty}\Gamma/{\eta_0^2R_0}$. In the plot here, the higher $V_E$ solutions are also the higher extensibility solutions, including the highest $M_w$ PEO solutions that lie in the upper right quadrant, and a clear distinction between less and more stringy solutions occurs as the high $V_E$ regime in the upper right corner is approached. 

The shear viscosity of dilute aqueous polymer solutions is relatively low, leading to small Oh for this geometry and hence IC pinching is observed before the transition to the EC regime. Rescaling $t_{tvec}$ and $\lambda_E$ by the IC time, $t_{ic} = \sqrt{\rho R_0^3/\Gamma}$ in Figure 6c results in two new dimensionless values: $t_{tvec}/t_{ic}$, and $\lambda_E/t_{ic}$. The two ratios appear to substitute of $\eta_E^{\infty}$ for the $\eta_0$ and $\lambda_E$ for $\lambda$ respectively in two well-known dimensionless groups: Ohnesorge number $Oh=t_{vc}/t_{ic}$, and Deborah number, $De=\lambda_E/t_{ic}$. Therefore, we christen these new ratios as the extensional Ohnesorge number $Oh_E =t_{tvec}/t_{ic}$, and the extensional Deborah number $De_E=\lambda_E/t_{vc}$. As the $Oh_E$ is a product of $Oh$ and $Tr^{\infty}$,and $De_E$ a product of $\Lambda$ and $De$, the difference in extensibility and flexibility is accentuated in the extensional $Oh-De$ plot. Furthermore, if we define extensional stringiness number as $Oh_EDe_E=S_E$, the increase in stringiness points in the direction shown in Figure 6d. Table 2 lists a representative subset of the timescales and dimensionless numbers including $S_E$ calculated for the polymer solutions represented in Figure 6 (see supporting information for the complete table). By plotting extensional Ohnesorge and extensional Deborah numbers and inferring extensional stringiness $S_E$ from their product, we highlight the magnified role played by finite extensibility of macromolecules. On explicitly computing the extensional stringiness, we find $S_E=\eta^{\infty}_E \lambda_E/\rho R_0^2$. Considering that an elasticity number $El =\eta_0 \lambda/\rho R_0^2$ defined using shear relaxation time and zero shear viscosity captures the elastic properties of nearly unperturbed polymer chains (and is thus based on linear viscoelastic response), the extensional stringiness $S_E=El_E$ can be thought of as a stretched elasticity number, that captures the profound influence of highly stretched chains and nonlinear viscoelasticity. Extensional stringiness, $S_E$ or stretched elasticity, $El_E$ number does not include any interfacial tension term, whereas viscoelastocapillary extensional number, $V_E$ depends on surface tension (and on elastocapillary number). Lastly, it can be shown that $V_E=De_EOh_EOh^2$ and therefore $V_E(DeOh)_E=S_E^2Oh^2$ or $V_E=S_EOh^2$, leading us to think of imaging-based classification of stringiness as $V_EDeOh$ stringiness!

The present study plus the datasets included herein, represent only a subset of the polymers used as rheological modifiers in formulations where control over stringiness is desirable, and comparisons are made only for unentangled solutions. Many formulations incorporate polymers at higher concentrations, such that the response of topological interactions or entanglements to deformation and flow have an influence. Additionally, many formulations involve polymers that can exhibit supramolecular interactions such as electrostatic interactions in polyelectrolytes, hydrophobic sticker-sticker association in telechelic and multiarm associative polymers \cite{jimenez_extensional_2018, jimenez_capillary_2020, martinez_narvaez_dynamics_2021}. We observe that stringiness may be enhanced through both an increase in extensional viscosity and extensional relaxation time. Thus even though increasing viscosity may increase the potential for stringy behavior, a high viscosity may not be conducive for the applications for which these formulations are made (i.e. spraying, pumping and dispensing, etc). In formulations where stringiness is desirable while maintaining a low viscosity, the appropriate polymer additive may provide a more suitable solution. Although shear viscosity alone cannot be a determination of stringiness, it still plays an important role in formulation design. For sprays and fiber spinning techniques, increased viscosity may pose problems when atomizing sprays due to the necessity for large backpressure and the rise of poor drop uniformity. Thus, considering the utilization of high $L_E$ with higher molecular weight, effectively enhances stringiness and formulation performance while reducing any negative effects high viscosity may cause. In contrast, food systems utilize higher viscosity for dips and sauces, like ketchup and mayonnaise. The higher viscosity aids in the stabilization of ketchup and mayonnaise, while at the same time, allowing for the sauce to be dispensed, spread, and "stuck" onto food. An increase in viscosity still prolongs the filament lifespan of an unstable liquid bridge to an extent. Paired with the elasticity of proteins and polymer-like ingredients in these food products, stringiness may be increased even further. While the focus systems of this current study has focused on unentangled polymer solutions, data from multiple publications, with polymer systems with varied macromolecular properties, support a variety of fields in which stringy formulation design is critical. While measures of stringiness have been carried out in previous studies using various different parameter comparisons, a general mapping of stringiness has never been posed in this manner.

\section*{Conclusions}
Stringiness of polymer solutions with a range of macromolecular properties was evaluated using dripping-onto-substrate (DoS) and dripping-into-air (DiA) protocols. A comprehensive analysis of experiments and past studies reveals that polymer solutions are often more stringy than Newtonian fluids of matched viscosity, surface tension, and density, and for the same polymer, increasing concentration or molecular weight leads to enhanced stringiness. In DoS protocols, stringiness can be envisioned as a consequence of longer filament lifespan and slower pinching dynamics. In DiA or dripping, the filament length, lifespan, and thinning rate appear to play a role. Here we show that even if the shear viscosity of polymer solutions is comparable, both DoS and DiA studies reveal dramatic differences in filament lifespan, extensional relaxation time and steady, terminal extensional viscosity. For these viscoelastic fluids, the filament lifespan is set primarily by extensional rheology response measured in terms of extensional relaxation time or terminal extensional viscosity. A nomograph with the two axes of $\eta_E^\infty$ and $\lambda_E$ provides a quick and representative assessment of stringiness. The polymer solutions explored in this study were all low viscosity fluids such that only inertiocapillary pinching is expected for Newtonian fluids with comparable shear viscosity. Therefore, we can construct a dimensionless version of the nomograph that we call the strongly extended OhDe graph, with $OhTr^\infty$ and $De\Lambda$ as the two axis. By incorporating considerations such as that of measuring extensional rheological properties and considering filament lifespan and filament length in addition to the typical shear rheological characterization, a more holistic approach to formulation design can be carried out. For applications such as agricultural sprays where drone-assisted rotary atomization is involved, polymer additives may be chosen more carefully to ensure that spray formulations are neither too viscous to atomize nor so dilute that spray drift occurs. In this study,we elucidated the effects of polymer additives in concentrations that fall into the \textit{unentangled regime}, and anticipate additional considerations are needed for entangled polymer solutions or melts, associative polymer solutions, and those systems that exhibit yield-stress behavior.

\section*{Author contributions}
All experiments described herein were conducted and analyzed by LE under the supervision of VS, and the protocols and research plan was developed and executed through collaborative discussions with the industrial partners

\section*{Conflicts of interest}
The authors report no conflict of interest.

\section*{Data availability}

The original data presented in this study is available on request from the corresponding author.

\section*{Acknowledgements}

We acknowledge the funding support from BASF industries and discussions with Dr. Prabodh  Varanasi and Dr. Rohini Gupta in particular. We also acknowledge discussions with Dr. Jelena Dinic (USG), Dr. Carina Martinez (University of Chicago), and Dr. Marc-Antoine Fardin (CNRS, France), as well as with the UIC ODES-lab students, especially Nadia Nikolova, and Abdullah Amer.

\section*{References}
\bibliography{references}

\end{document}